\documentclass{aastex62}

\submitjournal{ApJ}

\shorttitle{Magnetic Reconnection Invoked by Fast-Mode Shock}
\shortauthors{Zhou et al.}

\begin{document}

\title{Magnetic Reconnection Invoked by Sweeping of the CME-Driven Fast-Mode Shock}

\correspondingauthor{Guiping Zhou}
\email{gpzhou@nao.cas.cn}

\author{Guiping Zhou}
\affiliation{CAS Key Laboratory of Solar Activity, National Astronomical Observatories, Chinese Academy of Scienc, Beijing 100101, China}
\affiliation{University of Chinese Academy of Sciences, Beijing 100049, China}

\author{Guannan Gao}
\affiliation{Yunnan Observatories, Chinese Academy of Sciences, Kunming, Yunnan 650216, China}

\author{Jingxiu Wang}
\affiliation{CAS Key Laboratory of Solar Activity, National Astronomical Observatories, Chinese Academy of Scienc, Beijing 100101, China}
\affiliation{University of Chinese Academy of Sciences, Beijing 100049, China}

\author{Jun Lin}
\affiliation{University of Chinese Academy of Sciences, Beijing 100049, China}
\affiliation{Yunnan Observatories, Chinese Academy of Sciences, Kunming, Yunnan 650216, China}
\affiliation{Center for Astronomical Mega-Science, Chinese Academy of Sciences, Beijing 100012, China}

\author{Y. N. Su}
\affiliation{Key Laboratory of Dark Matter and Space Astronomy, Purple Mountain Observatory, Chinese Academy of Sciences, Nanjing 210023, China}
\affiliation{School of Astronomy and Space Science, University of Science and Technology of China, Hefei, Anhui 230026, China}

\author{Chunlan Jin}
\affiliation{CAS Key Laboratory of Solar Activity, National Astronomical Observatories, Chinese Academy of Scienc, Beijing 100101, China}
\affiliation{University of Chinese Academy of Sciences, Beijing 100049, China}

\author{Yuzong Zhang}
\affiliation{CAS Key Laboratory of Solar Activity, National Astronomical Observatories, Chinese Academy of Scienc, Beijing 100101, China}



\begin{abstract}

Coronal waves exist ubiquitously in the solar atmosphere. They are important not only in their own rich physics but also essential candidates of triggering magnetic eruptions in the remote. However, the later mechanism has never been directly confirmed. By revisiting the successive eruptions on 2012 March 7, fast-mode shocks are identified to account for the X5.4 flare-related EUV wave with a velocity of 550 km s$^{-1}$, and appeared faster than 2060$\pm$270 km s$^{-1}$ at the front of the corresponding coronal mass ejection in the slow-rising phase. They not only propagated much faster than the local Alf\'ven speed of about 260 km s$^{-1}$, but also simultaneously accompanied by type II radio burst, i.e., a typical feature of shock wave. The observations show that the shock wave disturbs the coronal loops C1 connecting active regions (ARs) 11429 and 11430, which is neighboring a null point region. Following a 40-min-oscillation, an external magnetic reconnection (EMR) occurred in the null point region. About 10 min later, a large-scale magnetic flux rope (MFR) overlaid by the C1 became unstable and erupted quickly. It is thought that the fast-mode shock triggered EMR in the null point region and caused the subsequent eruptions. This scenario is observed directly for the first time, and provides new hint for understanding the physics of solar activities and eruptions.

\end{abstract}



\section{Introduction} \label{sec:intro}

Coronal waves, as a common physical phenomena, mostly mean disturbance triggered by related impulsive energy releasing process, and exist ubiquitously in the solar corona. With largely improved observations, e.g., the global coverage and high time cadence of the Atmospheric Imaging Assembly \citep[AIA,][]{2012SoPh..275...17L} aboard Solar Dynamics Observatory (SDO) \citep[]{2012SoPh..275....3P}, coronal wave has been continuously understood not only in the aspects of the mechanisms of themselves, but also interests in whether such large-scale traveling disturbances can trigger remote eruptions \citep[e.g.,][]{2015LRSP...12....3W}).

Various waves invoked by the eruption in the solar atmosphere have been observed in different wavelengths and frequently reported. Among them, Extreme UltraViolet (EUV) wave has been studied most frequently owing to extensive observations from Solar and Heliospheric Observatory \citep[SOHO,][]{1995SSRv...72...81D}, the Solar Terrestrial Relations Observatory (STEREO), and especially SDO/AIA with high-cadence. EUV wave is often found to have two distinct parts: a leading wave-like front, and a slower trailing front ahead of the coronal dimming area \citep{1998GeoRL..25.2465T, 2014SoPh..289.3233L}. The leading front was interpreted as a fast-mode wave \citep[e.g.,][]{2001ApJ...560L.105W, 2008ApJ...681L.113V, 2009ApJ...700L.182P, 2011ApJ...738..160M}, and can propagate to the remote corona. In white light (WL) images, the fast EUV wave is often co-located with the leading shock-front of the related coronal mass ejection (CME) and is followed by the sharp density enhancements of CME structure \citep[e.g.,][]{2003ApJ...598.1392V}. In a composite of the WL and the EUV images, it is not difficult to recognize whether the EUV wave is the counterpart of a WL shock in the EUV band \citep[e.g.,][]{2007A&A...473..951T}. The impulsive cavity expansion during a coronal mass ejection (CME) may trigger fast EUV waves, most probably due to a fast lateral expansion \citep{2010A&A...522A.100P, 2008ApJ...681L.113V, 2018ApJ...868..107V}. A laterally propagating wavefront-like feature can be identified in STEREO/COR1 as the upper coronal counterpart of an EUV wave \citep[e.g.][]{2013ApJ...776...55K,2019ApJ...878..106H}. The consistent observations of coronal wave in EUV and WL observations help understand the physics of such propagating disturbance in different wavelengths, especially for the events observed from the solar limb.

On the aspect of determining modes of fast coronal waves related to flare/CMEs, in addition to the direct observational characteristics of actual velocity, other interferometric imaging in the radio domain, such as metric type II radio bursts, is a typical criterion for identifying fast-mode shocks as well. Type II radio bursts as reported for the first time by \citet{1947Natur.160..256P} and \citet{1950AuSRA...3..387W} appear as slowly drifting bands of emission in dynamic radio spectra. They are characteristic signatures of expanding coronal shock waves traveling outwards through the corona at speeds exceeding the local magneto-acoustic wave speed (or the local Alfv\'{e}n speed in the force-free environment) \citep[e.g.,][]{1960PASJ...12..376U}. The relationship between EUV wave and type II burst has soon been paid attention. \citet{2002ApJ...569.1009B} checked 173 EUV waves from 1997 March through 1998 June, and found only 23\% of EUV waves are accompanied by type II bursts. A consistent association rate of 22\% was obtained by \citet{2014SoPh..289.4563M} via analyzing 60 strong large-scale EUV wave events that occurred during 2007 January to 2011 February observed by the STEREO twin spacecraft. \citet[e.g.,][]{2009SoPh..259..227G} suggested that EUV waves may usually coincide with the occurrence of the metric type II burst in the early phase of eruptions. On the converse, \citet{2000A&AS..141..357K} studied 21 metric type II events observed in 1997 with the Potsdam radio spectral polarimeter, and noted that 90\% of metric type II bursts are associated with the EUV wave. Type II burst can therefore help identify the EUV wave with its correlated shock. However, only a fraction of the EUV waves (about 20-25\%) are indeed shock waves. The other fraction should be large-amplitude waves, but not shocks.

As a propagating disturbance triggered in the corona, coronal wave is usually questioned whether it may cause the eruption in a remote region. On one hand, we often observe that a proceeding flare triggers a sympathetic flare in distance \citep[e.g.,][]{2000JASTP..62.1449B}, or apparently activate distant filaments \citep[e.g.,][]{1949ApJ...110..382D}, or cause oscillation of a remote filament as observed in He I 10830 \AA~\citep{2008ApJ...685..629G}, and in EUV band with a transverse mode \citep{2011A&A...531A..53H, 2012ApJ...761..103G}. On the other hand, MHD oscillations in different modes are also observed during flaring process, e.g., global sausage mode \citep{2003A&A...412L...7N}, or kink mode in a magnetically-linked large loop \citep{2005A&A...440L..59F}, or Quasi-periodic pulsations \citep[QPP,][]{2005SSRv..121..115N}, and or a fast-mode shock \citep{2006ApJ...639L..99H, 2008ApJ...685..622A}. Among them, QPP is suggested as an observational evidence of the magnetic reconnection process \citep{2007ApJ...671..964T}, and also appears in stellar flares with similar manifestations \citep[e.g.,][]{2003A&A...403.1101M}. The close correlation of the coronal waves to flares/CMEs, indicate that, in addition to being the product of a nearby eruption, such fast propagating disturbance could also be the trigger of a remote eruption, e.g., exciting the sympathetic solar activity.

Excitation of different activities by the MHD wave is often investigated via numerical experiments. \citet{1982AZh....59..563Z} suggested that the flaring QPP can be modulated based upon sausage oscillations. In the work of \citet{ 2001ApJ...562L.103A}, QPP is invoked by the fast wave periodically modulating the distance between the slow shocks at the magnetic reconnection site. A current-carrying active region (AR) was suggested to be destabilized by the impact of a coronal wave \citep[e.g.,][]{2002ApJ...574..440O}. \citet{2005A&A...435..313M} noticed that the electric current density can greatly increase in the vicinity of a null point as triggered by the interaction of an aperiodic fast magnetoacoustic pulse with neutral points. A 2D simulation was performed by \citet{ 2006A&A...452..343N} to further find that fast MHD oscillations can periodically trigger magnetic reconnection by modulating periodical current in the null point region, and produce QPP when it transports in loops nearby. Based on the catastrophe model, \citet{2019MNRAS.490.2918X} demonstrated that different disturbances like a fast-mode shock can be produced by solar eruptions, and would be manifested as distinct EUV waves. EUV waves are mostly understood physically as fast-mode large-amplitude waves or shocks that are initially driven by the lateral impulsive expansion of a CME \citep[e.g.,][]{2008SoPh..253..249P, 2008ApJ...681L.113V, 2018ApJ...868..107V, 2010A&A...522A.100P, 2017SoPh..292....7L, 2012ApJ...750..134D}. One expectation to be understood is to focus on the probable contribution of coronal wave in triggering sympathetic solar activity. According to three-dimensional MHD model, several magnetic flux ropes were simulated to sympathetically erupt in a multipolar configuration by \citet{2011ApJ...739L..63T}, in which the main interacting mechanisms were attributed to magnetic reconnection and MHD instability. It is time to explore in details all the probable reasons causing magnetic reconnections one after another.  Fast coronal waves are frequently produced during solar eruptions, which may be one of the probable candidates leading to sympathetic activity and need to be paid attention especially from observations.

Previously, \citet{2019ApJ...873...23Z} studied a set of successive eruptive events on 2012 March 7, they laid emphasis on identifying a large-scale magnetic flux rope (MFR) connecting ARs 11429 to 11430, and a proceeding external magnetic reconnection (EMR) leading to the eruption of the MFR. In the present work, the same set of eruptions are re-visited to explore how the key EMR is triggered, and what the causal physics between the first and second sympathetic eruptions is. A fast coronal wave invoked by the first eruption is found as a new candidate to trigger the EMR and the later flare/CME. In \S\ref{sec:obs}, we briefly introduce observations and present results deduced from the observations which is followed by discussions and conclusions in \S\ref{sec:discussion}.

\section{Observations and results} \label{sec:obs}

The analysis is based on observations from SDO, STEREO, and SOHO as well as the radio data from the Hiraiso Radio Spectrograph \citep[HiRAS,][]{1995JCRL...42..111K} and Wind/WAVES \citep{1995SSRv...71..231B}. EUV data from SDO/AIA provide detailed information in six wavebands with the typical cadences of 12 s and the pixel sizes of 0.6 arcsec. The Helioseismic and Magnetic Imager \citep[HMI,][]{2014SoPh..289.3483H} aboard SDO gives the strength of the photospheric magnetic field with the time cadences of 45 s and pixel size of 0.5 arcsec. The solar rotation effect on all SDO data is removed through registering to 00:00 UT on 2012 March 07. STEREO-A (STA) and STEREO-B (STB) well record the propagating coronal wave ahead of the related CMEs from the side with trivial projection effects. In addition, coronal field strength is estimated based on the 3D coronal magnetic field model constructed by the flux rope insertion method \citep{2004ApJ...612..519V, 2009ApJ...691..105S, 2011ApJ...734...53S}.

\subsection{A summarization of EMR and 3 MFRs identified during the eruptions on 2012 March 7}

\begin{figure}[ht!]
\plotone{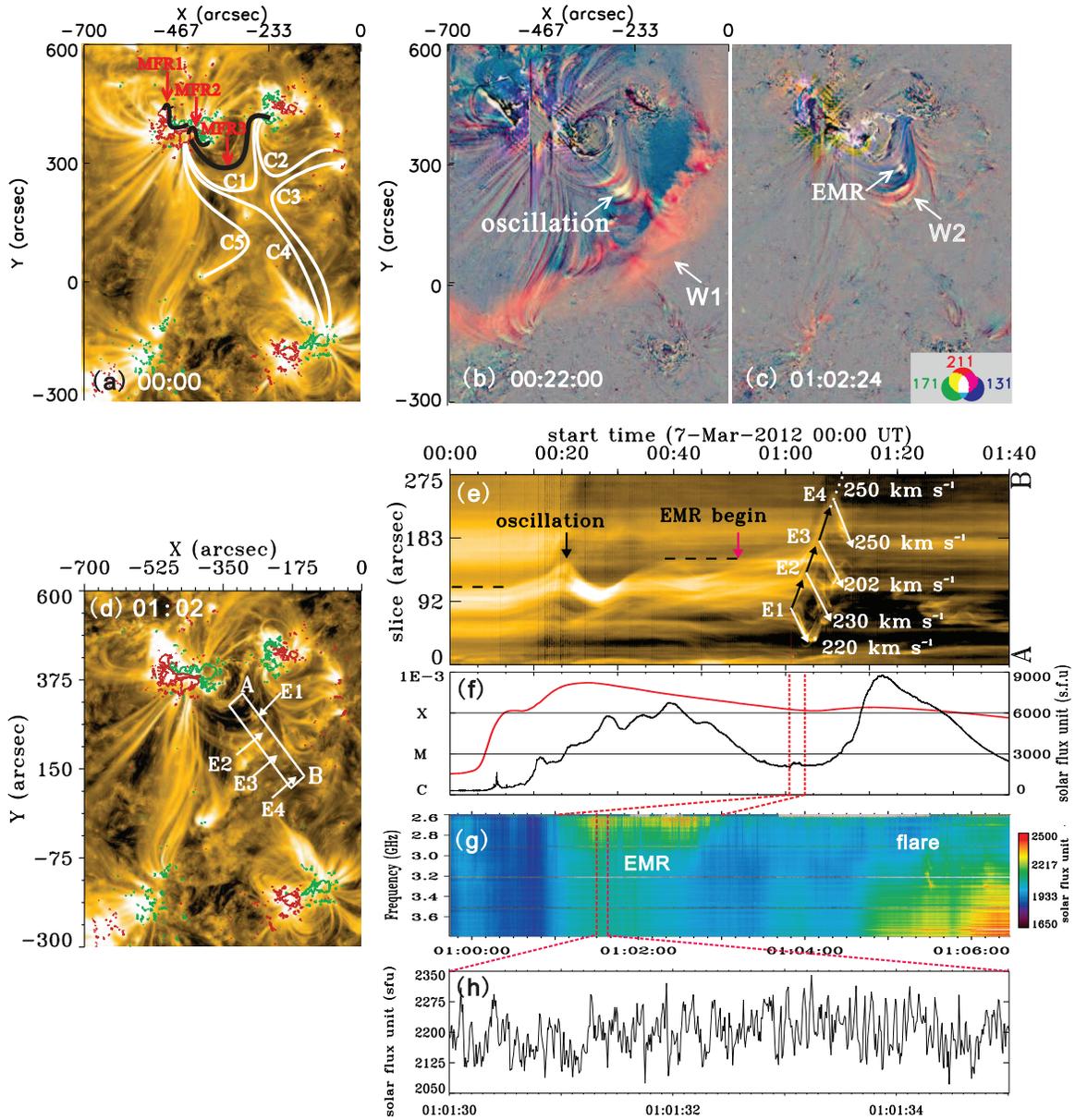}
\caption{Summary of three MFRs, EMR, and the associated null point region identified by \citet{2019ApJ...873...23Z}. Panel (a) EUV image in 171~\AA that shows MFRs 1 to 3 marked by black curves with the white curves highlighting the coronal loops C1 to C5. Panels (b) $-$ (c) Composites of running-difference images in three wavelengths of 131~\AA, 171~\AA, and 211~\AA, in which W1 originated from AR 11429 and caused C1 to oscillate for about 40 min, then EMR occurred and W2 was produced near the top of C1. The EMR process is displayed Panels (d) to (h). Panel (e) Time-sequence of the brightness distribution in EUV 171~\AA along the slice AB in Panel (d) to show the oscillation of C1 and four EMRs (E1 $-$ E4), among which E1–E3 produced bidirectional flows with speeds $> 200$~km~s$^{-1}$. Panel (f) Time profiles of the GOES X-ray flux in $1-8$~\AA (red) and the radio flux (black). E1 is observed to manifest QPP property in the radio spectra (Panels g to h). \label{fig:f0}}
\end{figure}

Time sequence of eruptions on 2012 March 7 are summarized in Table 1. Figure \ref{fig:f0} shows the identifications of three magnetic flux ropes (MFRs) and the external magnetic reconnection (EMR) as discussed in \citet{2019ApJ...873...23Z}. Here EMR describes the magnetic reconnection process that takes place high between a magnetic arcade and the magnetic fields outside this arcade \citep[e.g., ][]{2016ApJ...820L..37C, 2017ApJ...851L...1Z, 2019ApJ...878...46Z, 2020A&A...640A.101H}. The panels of Figure \ref{fig:f0} are mainly extracted from Figures 1 and 2 of \citet{2019ApJ...873...23Z}. In the EUV image in 171 \AA~of Figure \ref{fig:f0}a, three MFRs, namely MFR 1 to 3 are  marked by three black curves, and the coronal loops C1 to C5 are highlighted by the white curves to constitute a null point region. Figure \ref{fig:f0}b to \ref{fig:f0}c are composites of running-difference images in three wavelengths of 131~\AA, 171~\AA, and 211~\AA~ to display the EUV waves W1 and W2 manifested as red propagating fronts. The EMR is displayed in Figure \ref{fig:f0}d to \ref{fig:f0}h (see also Figure 2 of \citet{2019ApJ...873...23Z}). Figure \ref{fig:f0}e shows the time-sequence of the brightness distribution in EUV 171 \AA~along the slice AB in Figure \ref{fig:f0}d. W1 made C1 oscillate about 40 min before four EMRs (E1 $-$E4). E1 to E3 produced obvious bidirectional flows with linearly fitting speeds $>$200 km s$^{-1}$. Figure \ref{fig:f0}f presents the time profiles of GOES X-ray flux in 1$-$8~\AA~(red) and radio flux (black) to show EMRs occurring before the X1.3 flare. According to the radio spectrum (Figure \ref{fig:f0}g to \ref{fig:f0}h), E1 is also detected with the manifestation of QPP property in radio observation. The current work would lay emphasis on identifying the causal physics of EMR, and a new role of fast-mode shock on triggering sympathetic eruption. The corresponding analysis would be presented in the following sections.

\begin{deluxetable*}{lll}
\tablenum{1}
\tablecaption{Time sequence of the eruptions on 2012 March 7\label{tab:timeline}}
\tablewidth{0pt}
\tablehead{
\colhead{time} & \colhead{observations} & \colhead{explanation}
}
\startdata
00:02$-$00:40 UT & An X5.4 flare occurred in AR11429 with              & The X5.4 flare/CME is associated with the eruption of the\\
                         & a peak at 00:24 UT                                               & first MFR located in the north PIL of AR 11429. \\
00:15 UT           & The first CME related to the X5.4 flare first           & \\
                         & appeared in the field of view of STB cor1  & \\
00:10$-$00:35 UT & W1 oscillated coronal loops C1           & W1 was produced by the X5.4 flare/CME. C1 was disturbed\\
                         & connecting ARs 11429 and 11430, and lifted         & by the W1 to rise up. \\
                         & it up $>$25Mm & \\
00:52$-$01:06 UT & Small-scale eruptions occurred at the top of C1    & It is identified as EMR process that occurred in the null \\
                         & manifested by bidirectional EUV outflow and the   & region near the top of C1, which is indicated by the converging \\
                         & character of radio spectrum QPP                          & interface of C1$-$C4 and extrapolations.\\
00:55 UT           & The second CME appeared in the field view of      & This CME is related to the eruption of a large-scale MFR \\
                         & STB cor1. It originated near the top of C1     & connecting ARs 11429 and 111430, namely MFR3. \\
                         & with two flare ribbons between ARs 11429 and      & MFR 3 is well identified to have a temperature $\geq$10 MK. \\
                         & 11430, and propagated toward the earth.                  & Its two footpoints are situated in the negative and positive \\
                         &                                                                             & polarities from ARs 11429 and 11430, respectively, and have \\                                                                             &                                                                             & the same negative helicity but opposite current.\\
01:05$-$01:23 UT & An X1.3 flare occurred in AR 11429 with a peak    &\\
                         & at 01:14 UT                                                           & \\
01:25 UT           & The third fast CME related to the X1.3 flare first     & This CME was related to the eruption of the MFR 2, which \\
                         & appeared in the field of view of STB cor1    & is located in the south PIL of AR 11429.\\
\enddata
\tablecomments{EMR$-$external magnetic reconnection; C1$-$coronal loops connecting ARs 11429 and 11430; QPP$-$quasi-period pulsation; PIL$-$ploarity inversion line}
\end{deluxetable*}

\subsection{Fast EUV Waves Related to Successive Eruptions}

Three MFRs successively erupted on 2012 March 7 associated with three fast coronal waves and flares/CMEs. As identified by \citet{2019ApJ...873...23Z}, the second MFR is a large-scale structure connecting ARs 11429 to 11430. It became unstable and erupted after the occurrence of an EMR at the top. Observations show that this EMR region is disturbed by a fast coronal wave produced by the eruption of the first X5.4 flare-associated MFR in AR 11429. It is believed that such fast coronal wave could be a candidate driver of the key EMR.

\begin{figure}[ht!]
\plotone{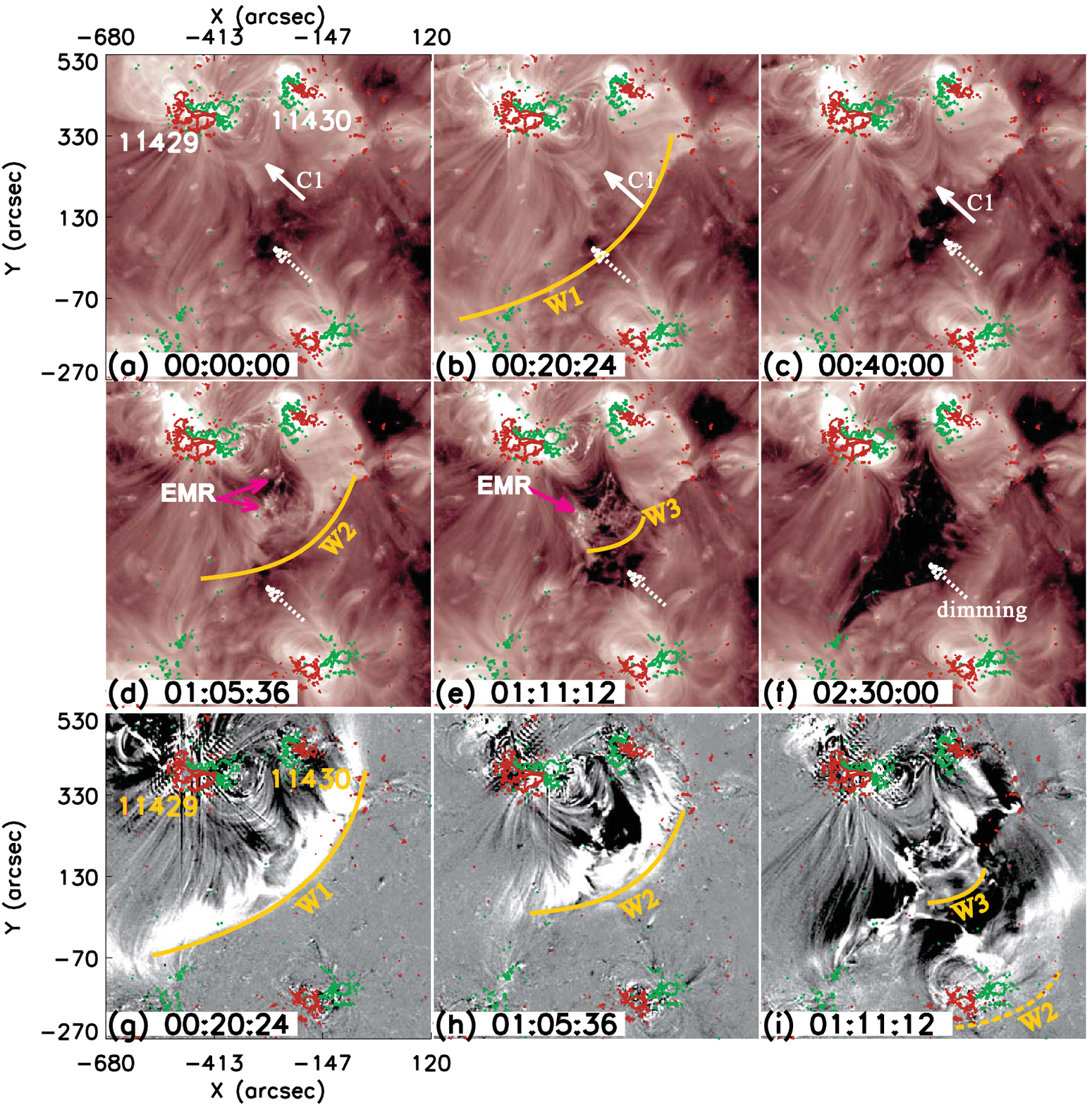}
\caption{W1$-$W3 related three fast successive coronal waves shown in the original ones (a$-$f) and the running-difference images (g$-$i) in 211 {\AA}. Yellow curves highlight the fronts of W1$-$W3. Arrows of solid white, dotted, and pink color denote the coronal loops connecting ARs 11429 and 11430, and the trans-equatorial coronal dimming, as well as the EMR, respectively. The corresponding animation is online. \label{fig:f1}}
\end{figure}

Three successive fast coronal waves were well observed in EUV passbands, especially in 211 {\AA} as illustrated in Figure \ref{fig:f1}. The original EUV images of Figures \ref{fig:f1}a to \ref{fig:f1}f well show coronal structures modified by eruptions. Three EUV waves are clearly seen as enhanced EUV density region with sharp fronts in the running-difference images of Figures \ref{fig:f1}g-\ref{fig:f1}i (see yellow arches as labeled by W1, W2, and W3) with difference time of 2 minutes. When the X5.4 class flare peaked at $~$00:20 UT, W1 was seen to sweep upward over the coronal loops connecting ARs 11429 to 11430 (see the white solid arrows and the label of C1), and raised C1 to a new height at the end of the X5.4 flare at 00:40 UT. At about 00:52 UT, the EMR (pink arrows in Figure \ref{fig:f1}d) intermittently started with the manifestations of small-scale EUV eruptions \citep[see also][]{2019ApJ...873...23Z}. In a short time, the second large-scale MFR erupted at about 01:00 UT to produce W2 as shown in Figure \ref{fig:f1}d. Immediately, the third MFR in the AR 11429 erupted leading to W3 with a clear appearance at about 01:11 UT (see yellow curve in Figure \ref{fig:f1}e). Eventually, a trans-equatorial coronal dimming was created (see the dashed arrow in Figures \ref{fig:f1}a$-$\ref{fig:f1}f). This suggests that the three successive eruptions are related to one another, and the fast wave caused by the prior eruption plays an essential role in triggering the eruption afterwards.

\begin{figure}[ht!]
\plotone{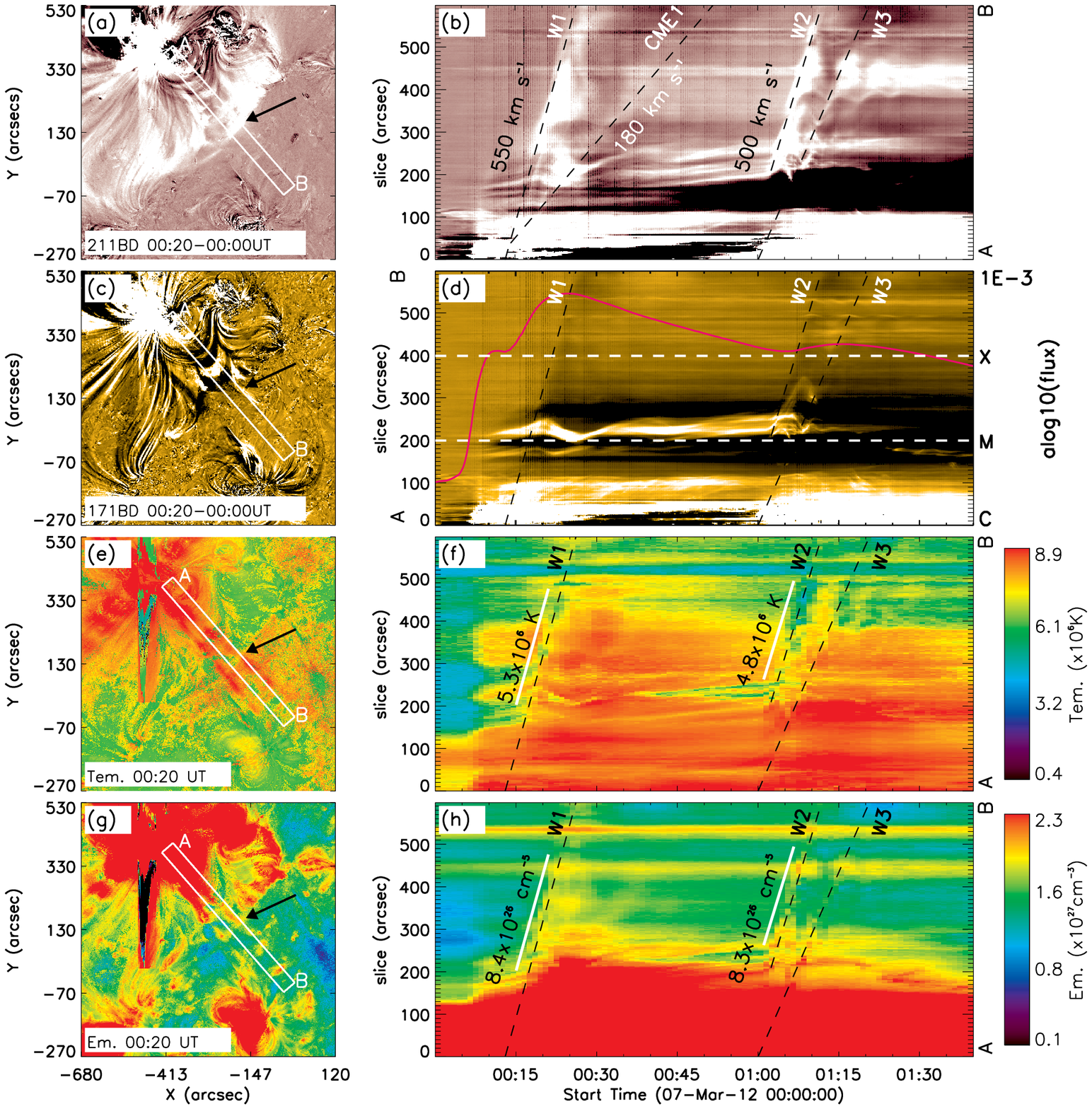}
\caption{Coronal waves W1$-$W2 shown in the time-slice stacking plots (right panels) along a narrow region AB in left panels. Top right two rows are the intensity distributions from base-difference data in 211 and 171 {\AA}, and bottom two ones are the temperature and density distributions deduced by DEM method with color bars on the right. The time profiles of GOES X-ray flux in 1$-$8 \AA (red) is overlaid on (d) with the coordinates on the right. By linearly fitting, coronal waves W1$-$W2 have speeds of about 550 and 500 km s$^{-1}$, respectively. W1-related CME1 structure has a slow velocity of about 180 km s$^{-1}$ lagging behind W1 projected in the sky plane. \label{fig:f2}}
\end{figure}

Physical properties of coronal waves, e.g. the speeds projected on the sky plane, the temperature and density of local corona, can be learned from the space-time map (right panels) over a narrow region AB (left panels) during the propagation of the wave noses as shown in Figure \ref{fig:f2}. What these panels display are, from the top to the bottom, the filtergrams in the base-difference data of 211~\AA and 171~\AA~by subtracting the one at 00:00 UT, distributions of the temperature and the density deduced via differential emission measure (DEM) approach as discussed in \citet{2012ApJ...761...62C}. The DEM code ``xrt\_dem\_iterative2.pro'' in SSW package was originally designed for the Hinode X-ray Telescope data \citep{2004ASPC..325..217G, 2004IAUS..223..321W}. It was then modified to compute the DEM based on the AIA data \citep[see also][]{ 2010ApJ...723.1180S, 2011ApJ...740....2W, 2012ApJ...761...62C}. The DN counts in each of the six EUV wavelength are divided by the exposure time and spatially averaged over all pixels in the area \citep[see][]{2012ApJ...761...62C}. The averaged count rates were used as the input of ``xrt\_dem\_iterative2.pro'' routine to calculate the DEM. The region DEM shows a broad temperature distribution from about 1.7 MK to 11 MK (logT =6.2$-$7.0). The total emission measure (EM) is calculated according to {\it EM}=$\int DEM(T)dT$ and is located in the range of [3.3$\times10^{26}$, 7.9$\times10^{29}$] cm$^{-5}$. Since the filling factor of the plasma is unknown, it is assumed to be 1 in the density calculation. An uncertainty of 20\% is considered as a lower limit for estimating T and EM. The right panels display the resultant space-time maps of the corresponding parameters during 00:00$-$01:40 UT. The right four rows of images are displayed in the range of [-20, 20] DN s$^{-1}$, [-90, 90] DN s$^{-1}$, [4.4$\times10^{5}$, 8.9$\times10^{6}$] K, [1.1$\times10^{26}$, 2.3$\times10^{27}$] cm$^{-5}$, respectively. The maximum temperature and emission measure for W1 are $5.6\times10^{6}$ K and $1.2\times10^{27}$ cm$^{-5}$. The coronal waves appear as bright moving feature in 211 {\AA} of Figure \ref{fig:f2}b, but dark ones in 171 {\AA} (Figure \ref{fig:f2}d). It can be interpreted as emission decrease in 171 {\AA} channel due to heating to higher temperatures as discussed in \citet{2015ApJ...812..173V}. By linearly fitting (see the dashed oblique lines), W1 and W2 are estimated to have speeds of $~$550 and $~$500 km s$^{-1}$ projected in the sky plane observed by SDO/AIA, respectively. The W1-related CME1 structure can also be seen as a propagating brightening structure in Figure \ref{fig:f2}b, and has a slower velocity of about 180 km s$^{-1}$ than that of the fast W1 ahead. These speeds give lower estimates, since we only observe the projected kinematics: EUV waves are propagating at a certain height above the curved solar surface \citep{2009ApJ...703L.118K, 2019ApJ...877...68P}, and the measured CME kinematics is projected against the plane-of-sky.

In order to determine the mode of the coronal waves, it is necessary to estimate the local Alfv$\acute{e}$n speeds {\it V$_{A}$} around C1 that was disturbed by the eruption. According to {\it V$_{A}$}={\it B}/$\sqrt{\mu_{0} \cdot \rho}$ $\approx$ $2.18\times10^{11}${\it B}/$\sqrt{n_{i}}$, with {\it B} being the magnetic field strength on C1 at the height of 0.7R$_{\bigodot}$ as detected by STEREO, which is about 3 G based on the magnetic field model constructed using the flux rope-inserting method \citep{2009ApJ...691..105S, 2011ApJ...734...53S}, and \emph{T} and $n_{i}$ being the associated average temperature and density of the local corona. According to \citet{1999ApJ...520..880A}, \emph{n$_{i}$}=$\sqrt{EM/w}$, with \emph{EM} being the emission measure, and \emph{w} being the width of C1 under the assumption of a circular cross-section \citep[see also][]{2016ApJ...823L..19Z}. We select the regions right ahead of wave fronts (see white solid bars) in the space-time maps in Figures \ref{fig:f2}f and \ref{fig:f2}h, and get \emph{T} of 5.3 MK and 4.8 MK, \emph{EM} of 8.4 $\times$ 10$^{26}$ cm$^{-5}$ and 8.3 $\times$ 10$^{26}$ cm$^{-5}$, respectively. Their mean density $n_{i}$ correspond to 6.2 $\times$ 10$^{8}$ cm$^{-3}$ and 6.1$\times$ 10$^{8}$ cm$^{-3}$ with the loop width \emph{w} measured as about 22 Mm. Therefore, W1 and W2 propagated in the corona with Alfv$\acute{e}$n speeds {\it V$_{A}$} of about 260 km s$^{-1}$. Considering a possible uncertainty of about 10\%$-$20\% in calculating the temperature, density, and width of the coronal loops, the local Alfv$\acute{e}$n speeds corresponding to W1 and W2 should not exceed 330 km s$^{-1}$. Consequently, as their speeds are faster than the local Alfv\'{e}n speed, we identify W1 and W2 as fast-mode shocks propagating through the corona.

\subsection{Fast-mode Shocks exist ahead of CME1 and CME2}

\begin{figure}[ht!]
\plotone{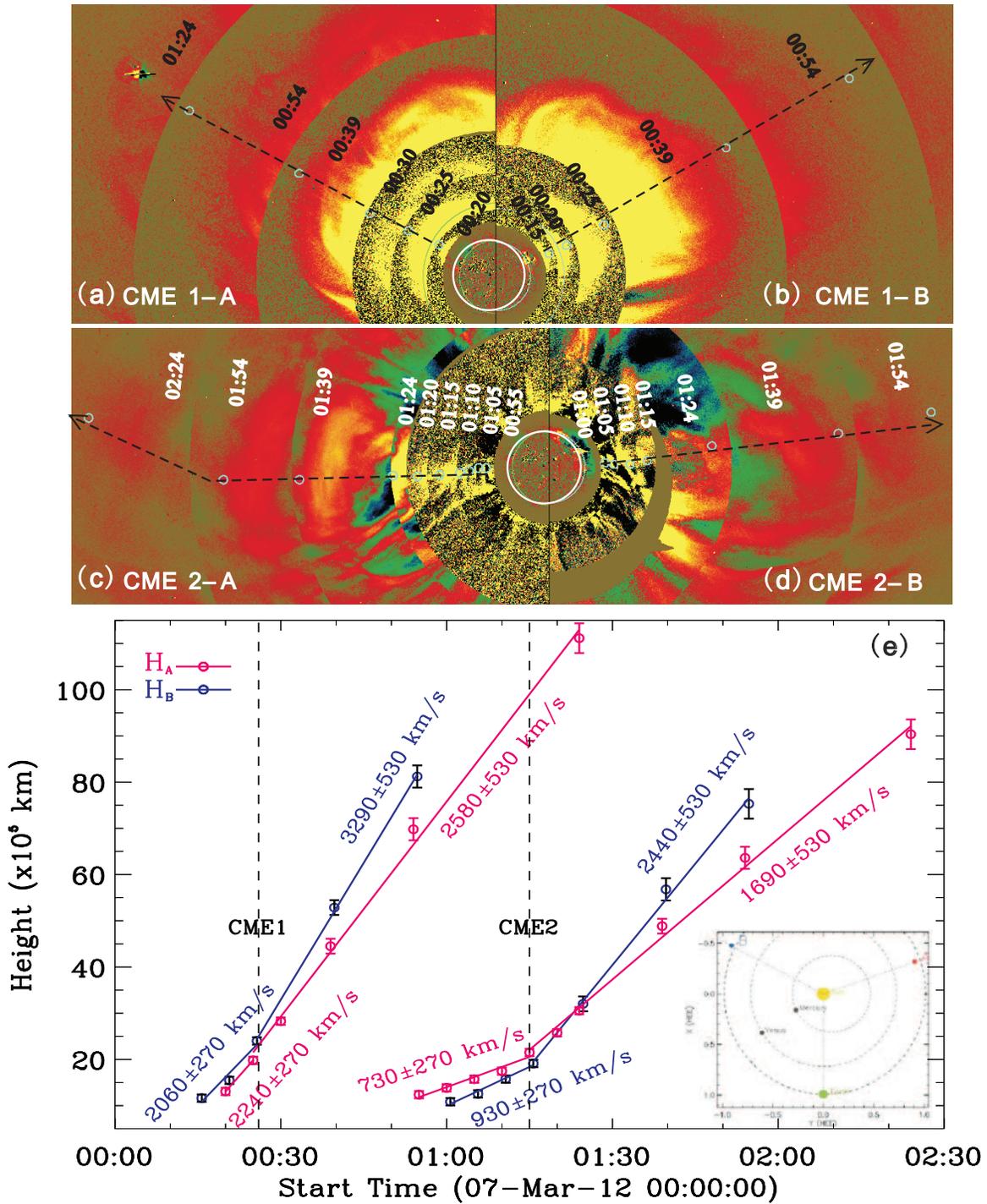}
\caption{CME 1 (top row) and CME 2 (middle row) in the composite of running-difference images from STA (left panels) and STB (right panels). Small circles connected by dashed lines denote the propagating fronts at different times. Panel (e) presents the corresponding variations of the height of each CME with red colors for STA and blue ones for STB. The locations of STA/STB, the earth/SDO and the Sun are displayed in the inset at the right bottom corner. \label{fig:f3}}
\end{figure}

Figure \ref{fig:f3} shows the first CME (CME1 hereafter) associated with W1 during 00:15$-$01:24 UT, and the second CME (CME2 hereafter) related to W2 during 00:55$-$02:24 UT observed by the COR1 and COR2 coronagraph on board STA (left panel) and STB (right panel) in the top two rows. In order to show the whole propagating process clearly, the composites of images obtained by STA and STB in which the CME front can be recognized at different moments are displayed in Figures \ref{fig:f3}a through \ref{fig:f3}d. Connecting every adjacent two forefronts as denoted by small circles via a dashed line indicates the straightforward propagation of CME1 and the apparent deflection of CME2 in the way of leaving the Sun. Figure \ref{fig:f3}e illustrates height-time variations of the two CME fronts. Here the height is measured from the solar center. Blue and red colors denotes observations from STA and STB, respectively. The spatial locations of STA/STB, the earth/SDO and the Sun are showed at the bottom right corner of Figure \ref{fig:f3}e.

Combining observations from STA with those from STB, we could deduce reasonable values for velocities of CME1 and CME2. The height-time distributions of CMEs in Figure \ref{fig:f3}e observed by STA (red) and STB (blue) clearly show two phases of a CME, i.e., the initial phase followed by the impulsive one, which are linearly fitted with solid lines, respectively. In order to estimate uncertainties of CME heights, we repeatedly trace CME fronts in the sky plane at each moment ten times to get 3$\sigma$ of the measured distances as the error bars. The errors of velocities are obtained from the Monte-Carlo simulations with the uncertainties of CME front situations. We use the maximum velocity uncertainty in different stages of a CME as the corresponding error bars. As a result, CME1 has an initial speed of 2060$\pm$270 km~s$^{-1}$ from 00:15 to 00:25 UT, and a later impulsive speed of 3290$\pm$530 km~s$^{-1}$ observed by STB, corresponding to 2240$\pm$270 km~s$^{-1}$ and 2580$\pm$530 km~s$^{-1}$ in the field of view of STA. With the similar fitting method, CME2 has an initial speed of 730$\pm$270 km~s$^{-1}$ from 00:53 to 01:15 UT, an impulsive speed of 1690$\pm$530 km~s$^{-1}$ from STA, which are 930$\pm$270 km~s$^{-1}$ and 2440$\pm$530 km~s$^{-1}$ from STB. Comparing with the local Alfv\'{e}n speed of a few 10$^{2}$~km~s$^{-1}$, we believe that the occurrence of the fast-mode shocks in front of both CMEs are inevitable.

\begin{figure}[ht!]
\plotone{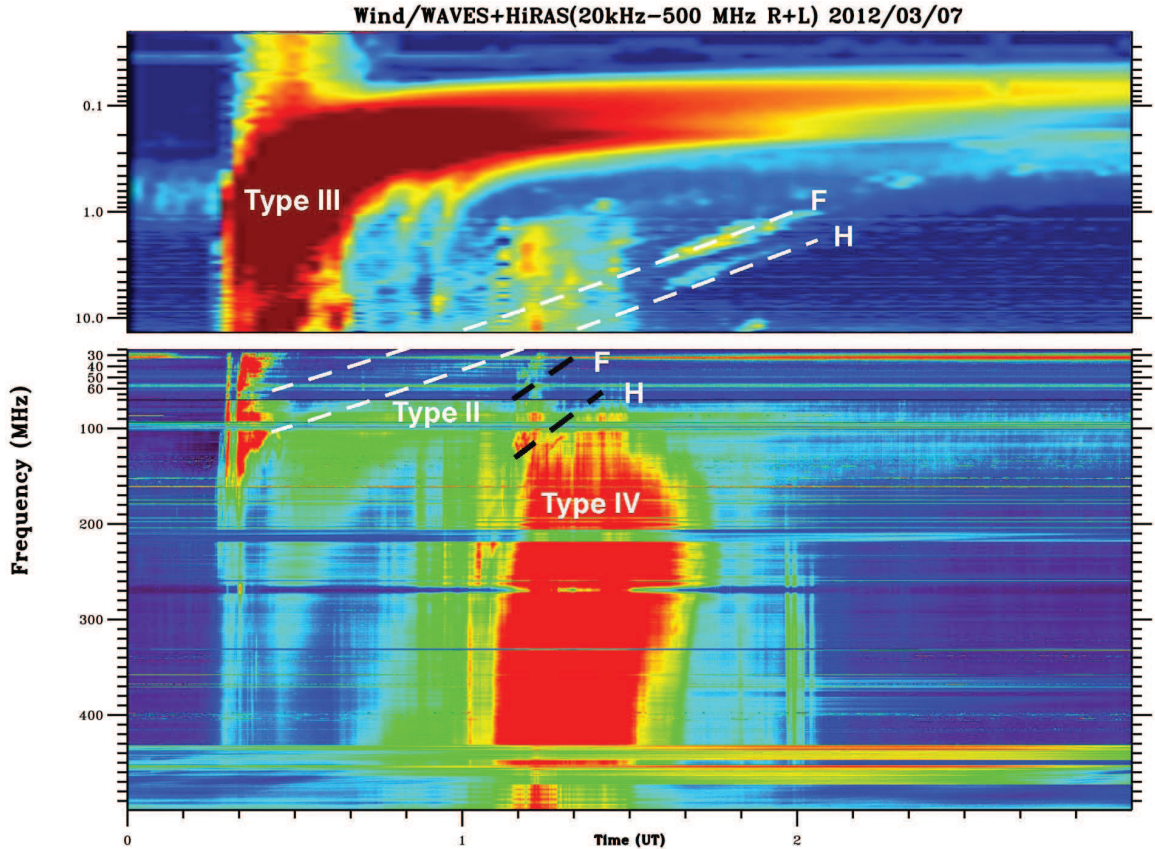}
\caption{Type II radio burst from radio spectrum observations accompanying with W1 and W2 (see white and dark dash curves), both of them display the fundamental (``F'') and harmonic (``H'') band ). \label{fig:f4}}
\end{figure}

Type-II radio bursts are usually considered as a good manifestation to identify coronal shock \citep[e.g.,][]{1960PASJ...12..376U}. Figure \ref{fig:f4} shows the radio dynamic spectrum in the frequency range from 20 kHz to 500 MHz obtained from Wind/Waves and HiRAS during 00:25-03:00 UT covering two times of eruptions. Two sets of type-II radio burst signals covering time intervals of 00:25-03:00 UT and 01:10-01:25 UT can be recognized (see white and black dashed curves). The initial frequencies of the fundamental components are located at 55 MHz and 75 MHz for the type-II radio bursts driven by the shocks associated with two eruptions, respectively. According to the empirical model about the coronal density distribution of \citet{1999ApJ...523..812S}, the density at the coronal base is considered as 10$^{10}$ cm$^{-3}$. The initial heights of the two shocks were deduced at about 0.9 R$_{\bigodot}$ and 0.8 R$_{\bigodot}$, respectively, which is consistent with the observations from STA/STB. Since the noses of W1 and W2 propagated in the direction along the slice AB (see the left panels of Figure \ref{fig:f2}), type-II radio bursts are suggested to relate to the eruptions along this way. \citet{2020arXiv200910872F} recently constructed the three-dimensional configurations of the wave W1 surfaces using three techniques. They suggested a similar result that the same W1 nose increased from the speeds 600 to 800 km s$^{-1}$ across the solar disk with the manifestation of EUV disturbances to a much higher speed up to 3800 km s$^{-1}$ in the extended corona. We discussed that W1 and W2 in EUV observations from SDO may be the footprints in the lower corona of the corresponding shocks driven by CME1 and CME2, which are the enhanced EUV emission at the wave front caused by the downward push and compression of the plasma at the base due to the coronal shock or large amplitude wave \citep[e.g.,][]{2005ApJ...626L.121W,2011ApJ...737L...4H, 2011ApJ...743L..10V}. In any case, the current work shows that shock waves were produced during the successive eruptions on 2012 March 7 whether in the lower corona or in the higher one. Combining the analysis of observations, a reasonable physical mechanism about the shock wave is deducted to trigger the key EMR causing a sympathetic eruption.

\section{Discussions and Conclusion} \label{sec:discussion}

\begin{figure}[ht!]
\plotone{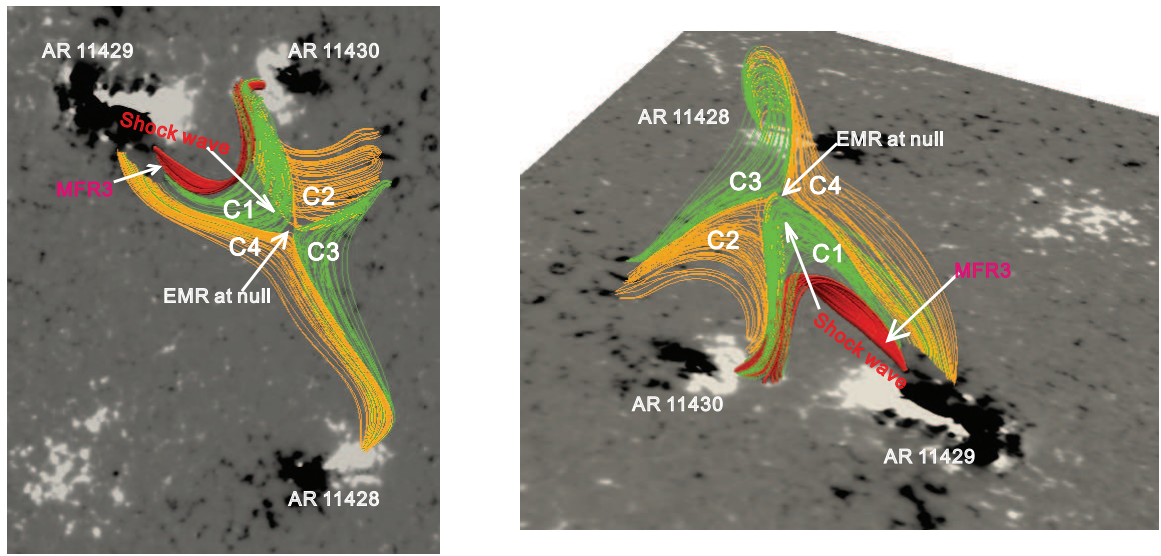}
\caption{3D magnetic topology constructed by Flux Insertion Method on a vertical magnetogram at 00:00 UT seen from the top (left panel) and side view (right panel). W1 disturbs the green coronal loops overlying the large-scale MFR of red color. The null point region is located at the junction of two ``X'' configurations of yellow and green field lines. \label{fig:f5}}
\end{figure}

In order to well understand how the X5.4 flare-related shock wave triggers magnetic reconnection, we need to check the 3D magnetic topology of the erupting structures at 00:00 UT. The 3D magnetic topology covering a trans-equatorial region was reproduced according to magnetic field modeling using the flux rope insertion method \citep{2004ApJ...612..519V, 2009ApJ...691..105S, 2011ApJ...734...53S}. A flux bundle is inserted with an initial axial flux of 4$\times$10$^{20}$ Mx and poloidal flux of 0 Mx cm$^{-1}$ in the best-fit model. It then gradually evolved into a flux rope with increasing poloidal flux due to magnetic reconnections with the surrounding fields during the magneto-frictional relaxation process as discussed in \citet{2011ApJ...734...53S}. After 30000-iteration relaxations, the initial inserted flux bundle was turned into a flux rope as shown in Figure \ref{fig:f5}. The size of the modeled high-resolution magnetic field region is about 59$^{\circ}$ in longitude and 80$^{\circ}$ in latitude on the solar disc up to a height of 1.25 R$_{\sun}$ above the solar surface, and the spatial resolution in the low corona is 0.002 R$_{\bigodot}$. Figure \ref{fig:f5} shows the 3D constructed configuration seen from the top (left) and from the side (right), which manifest topological features similar to those observed in EUV by SDO/AIA. Red lines are for the large-scale MFR3 connecting ARs 11429 to 11430, above which four sets of field lines (``C1-C4''; indicated in yellow and green colors) consist of two X-type configurations constituting the null point region.

The modeled 3D coronal magnetic configurations displayed in Figure \ref{fig:f5} show how the shock triggers the EMR. Driven by CME 1, the fast-mode shock associated with W1 (see also Figures \ref{fig:f1}g, \ref{fig:f2}a, and \ref{fig:f2}b) globally propagated with a strong front toward the loop top of C1 and making it disturb between ARs 11429 and 11430 for about 40 min since 00:15 UT. At about 00:52 UT, EMR occurred in the null point region around C1, manifested by small-scale EUV eruptions in FOV of SDO/AIA (see also Figures \ref{fig:f0}d, \ref{fig:f0}e, \ref{fig:f1}d, and \ref{fig:f1}e), and the typical QPP features (see Figures \ref{fig:f0}f through \ref{fig:f0}h) revealed by radio observations \citep[see][]{2008SoPh..253..117T,2019ApJ...873...23Z}. With the occurrence of the intermittent EMR, the force that results in the overlying magnetic field and keeps the configuration in the equilibrium is weakened, leading to the eruption of the large-scale MFR3 (see also Figure \ref{fig:f0}a) at about 01:03 UT. Our observations here show that W1-related fast shock was followed by the EMR, and the large MFR3 erupting as a fast halo CME. The physical process behind this scenario could be the accumulation of the electric current in the region near the null point, which is triggered periodically by the fast magneto-oscillation as W1 passed through the null point region. This is consistent with the simulation results of \citet{2005A&A...435..313M} and \citet{2006A&A...452..343N}. In addition, we also suggest that W1 might greatly increase the local turbulence in the null point region, giving rise to a large anomalous resistivity that eventually invoked EMR.

In  spirit of the works of \citet{2005A&A...435..313M} and \citet{2006A&A...452..343N}, we revisit the three successive eruptions taking place on 2012 March 7. Propagating characteristics of the CMEs in two eruptions indicate the occurrence of the fast-mode shock, which are confirmed by the type-II radio burst observed in the same time intervals when they were observed. By analyzing the initial altitudes of the two shocks, and looking into the magnetic configurations associated, a null point region was recognized, and the fast-mode shock driven the first CME passing through the region around the null point. According to \citet{2005A&A...435..313M} and \citet{2006A&A...452..343N}, what we observed in the present event might result from that the electric current intensity exceeds the threshold value, which leads a spontaneous magnetic reconnection to taking place in the null point region, and alternates the topological structure of the related magnetic configuration, yielding the consequent eruption eventually. This scenario seems to suggest that the fast-mode shock invoke an oscillation in the nearby magnetic structure including a null point, triggering the magnetic reconnection process at the null point region, and further the sympathetic eruption in the related magnetic configuration. More observations in the future to verify this scenario are surely needed.

\acknowledgments

The work is supported by the B-type Strategic Priority Program of the Chinese Academy of Sciences (Grant No. XDB41000000), the National Natural Science Foundation of China (11322329, 11533008, 11663007, 11903050, 11025315, 11221063, 11941003 and 11303049), Key Research Program of Frontier Sciences, CAS (Grant No. ZDBS-LY-SLH013), the Chinese Academy of Science Project KJCX2-EW-T07, the National Key Basic Research Science Foundation (G2011CB811403 and G2011CB811402), and Beijing Natural Science Foundation 1202022. The work of JL was supported by the Strategic Priority Research Program of CAS with grants XDA17040507 and QYZDJ-SSWSLH012, the NSFC under the grant 11933009, and the project of the Group for Innovation of Yunnan Province grant 2018HC023 and the Yunnan Province Yunling Scholar Project. YS is supported by the NSFC 41761134088, 11473071 and 11790302 (11790300).

%

\vspace{5mm}

\end{document}